\documentclass[11pt]{article}
\usepackage[utf8]{inputenc}
\usepackage{amsmath,amssymb}
\usepackage[margin=1in]{geometry}
\usepackage{setspace}
\usepackage{pdflscape} 
\usepackage{graphicx}

\onehalfspace

\usepackage{booktabs,tabularx,dcolumn} 
\usepackage[style=apa,backend=biber,sortcites=true]{biblatex}
\usepackage{multirow}

\usepackage{rotating}

\usepackage{hyperref}

\title{Discrete Exponential-Family Models for Multivariate Binary Outcomes}
\author{George G. Vega Yon\footnote{george dot vegayon at utah dot edu.}\and Mary Jo Pugh\and Thomas W. Valente}
\date{November 1, 2022}

\renewcommand{\Pr}[1]{{\mathbb{P}\left(#1\right) }}

\newcommand{\sufstats}[1]{s\left(#1\right)}
\renewcommand{\exp}[1]{\mbox{exp}\left\{ {#1} \right\}}
\renewcommand{\log}[1]{\mbox{log}\left\{ {#1} \right\}}
\newcommand{\transpose}[1]{{#1}^\mathbf{t}}

\newcommand{\s}[1]{\sufstats{#1}}

\renewcommand{\beta}{\theta}

\newcommand{\normconst}[1]{\kappa\left(#1\right)}

\graphicspath{{./figures/}{.}{./terms/}}

\usepackage{pstricks}

\definecolor{USCCardinal}{HTML}{990000} 
\definecolor{USCGold}{HTML}{FFCC00}
\definecolor{USCGray}{HTML}{CCCCCC}

\renewcommand{\Pr}[1]{{\mathbb{P}\left(#1\right) }}

\newcommand{\Prcond}[2]{{\mathbb{P}\left(#1{\vphantom  {#2}}\;\right|\left.{\vphantom  {#1}}#2\right)}}

\renewcommand{\exp}[1]{\mbox{exp}\left\{#1\right\}}

\newcommand{\logit}[1]{\mbox{logit}\left(#1\right)}
\newcommand{\logitinv}[1]{\mbox{Logit}^{-1}\left(#1\right)}

\newcommand{\snamed}[2]{\s{#1}_{\mbox{#2}}}

\newcommand{\tp}[1]{{#1}^{\mathbf{t}}}

\newcommand{\y}[1]{y_{{#1}}}
\newcommand{\yvec}[1]{\mathbf{\y{#1}}}
\newcommand{\ycal}[1]{\left\{0, 1\right\}^{#1}}

\newcommand{\sstat}[1]{s\left(#1\right)}

\newcommand{\chng}[1]{\delta\left(\y{#1}:0\to1\right)}

\newcommand{\linevec}[2]{\left(\begin{array}{ccc}#1&#2\end{array}\right)}

\renewcommand{\exp}[1]{\mbox{exp}\left\{#1\right\}}

\begin{document}

\graphicspath{{.}{fig/}}

\maketitle

\begin{abstract}
Studies that collect multi-outcome data such as tobacco and alcohol use are becoming increasingly common. In principle, multi-outcomes studies investigate the correlations between outcomes, including, causal links and/or joint distributions. Although there are many methods for studying multivariate outcomes, significant limitations regarding scale and interpretation persist. Here we introduce a model based on the exponential-family for discrete binary outcomes that  provides a flexible framework for hypothesis testing of multiple binary outcomes in a computationally efficient fashion.
\end{abstract}

\section{Introduction}

Complex systems are all around us, and we now have the tools to study them rightfully. In the last few decades, the increasing availability of data and computational power has allowed the scientific community to address complex questions using sophisticated and advanced statistical methods that were previously unattainable. More computing power and data allow one to take down common assumptions and study social and natural phenomena embracing their complexities. In particular, we can analyze complex systems without assuming independence and leverage their components' interdependence to better understand their nature.

\begin{figure*}
    \centering
    \begin{minipage}[m]{.45\linewidth}
    Possible ways to model outcomes $\{y_a,y_b\}_t$:
        \begin{align*}
        y_{t,a} &\sim f(X),\mbox{ or} \\ 
        y_{t,a} &\sim f(y_{t,b}, X),\mbox{ or} \\ 
        y_{t,b} &\sim f(y_{t,a}, X,\mbox{ or} \\ 
        y_{t,b} &\sim f(y_{t-1,b}, y_{t-1,a}, X),\mbox{ or} \\
        \dots
        \end{align*}
    \end{minipage}\hfill %
    \begin{minipage}[m]{.45\linewidth}
	\centering
	\includegraphics[width=.7\linewidth]{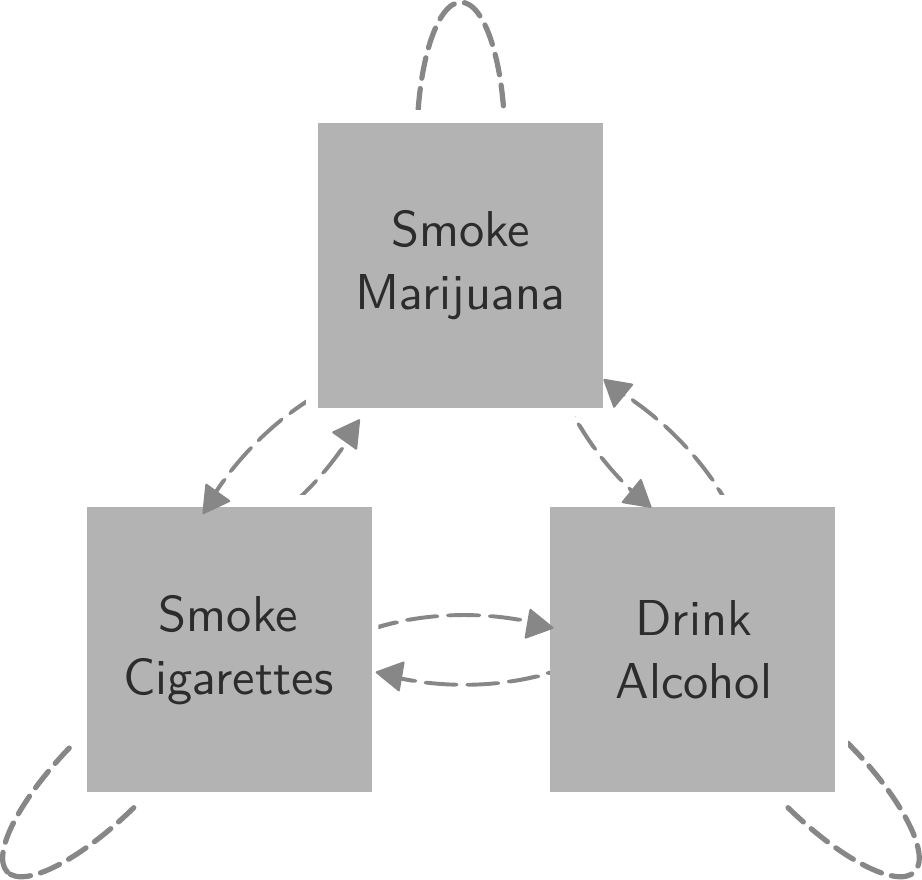}
    \end{minipage}
    \caption{When dealing with multiple outcomes, we tend to assume one as a function of the others. Yet, assessing the validity of such assumptions can be elusive; especially when the outcomes have multiple ways of interacting/mediating each other.}
    \label{fig:dags}
\end{figure*}

In social sciences, beyond social networks, many phenomena can be described as complex systems.  Individual preferences, attitudes, and behaviors are interrelated. For example, individuals who have preferences and attitudes that support a healthy lifestyle are less likely to engage in negative health behaviors such as tobacco use, excessive alcohol use, and other negative health behaviors. Someone who  drinks alcohol and consumes Marijuana may be more likely to also consume tobacco. We may also hypothesize that someone who suffers from depression is more likely to have sleep problems; and to add complexity to the phenomena, both depression and sleep problems may be involved in a feedback loop as shown in Figure \ref{fig:dags}. Consequently, developing models to accurately analyze multiple outcomes is prudent analytically and is reflective of population reality.

In this paper, we introduce a model from the exponential family designed to analyze multiple binary outcomes,  called the  Discrete Exponential-Family Model [DEFM] \parencite{VegaYonPhD2020}. DEFM is derived from the  large family of distributions known as exponential random graph models [ERGMs] used to analyze  social networks. DEFMs provide a computationally efficient way to analyze multi-outcome data and allows for the evaluation of complex hypotheses involving the interaction of multiple dependent dichotomous variables. The manuscript continues as follows: the next section presents a literature review of statistical models of multiple outcomes; the methods section introduces DEFMs formally, including parameter interpretation and goodness-of-fit evaluation; the empirical section presents an application using example data, and finalizes with a discussion including limitations and future research directions.

\section{Modeling Multiple Correlated Binary Outcomes}

Many problems across sciences deal with predicting or analyzing dichotomous data such as a yes/no response. The Logit and Probit regression models are widely used for such outcomes and are taught across scientific fields \parencite{Agresti2007}. Although Logit and Probit models are useful for dichotomous outcomes, they cannot be used to simultaneous analyze multiple such outcomes. As researchers acknowledge the inter-relatedness of multiple such outcomes, developing appropriate methods has become increasingly necessary \parencite{Martin2021}.

We can approach data analysis in two ways: separating them into dependent and independent variables, looking at one at a time, or using more complicated statistical methods that account for complex relations \parencite[see for example][]{Fujimoto2012}. The first approach, although practical, has a huge limitation: we can only set one of the outcomes as a dependent variable. For example, understanding the associations between independent variables and smoking and drinking requires analyzing smoking and drinking separately. This approach can lead to biased estimates, incorrect inferences, and/or inflated error rates \parencite{Davenport2018}. The second approach is the proper statistical treatment, for which several models have been developed. Nonetheless, many take the correlation as a problem rather than part of the analysis.

\subsection{Cross-sectional studies}

A recent paper \parencite{Bai2020} provides an extensive literature review for multi-outcome models using latent variables. Using the pairwise composite-likelihood method, they develop a model for multivariate mixed-type data based on latent continuous variables. \textcite{CAREY1993}  introduces a way to fit models with jointly distributed binary outcomes iterating between Generalized Estimating Equations [GEEs] and Logistic regression. As they indicate, GEEs are good when the correlations are not the model’s focus. They also point out that correlation structure estimates using GEEs can be seriously inefficient. When trying to estimate the second-order GEE, identifying the correlation structure, they point out that such a process is computationally infeasible.

\textcite{Davenport2018} presented a method mixing kernel regressions with GEEs, but it does not specify or identify the correlation structure, which is needed to understand how the outcomes are related. \textcite{TeixeiraPinto2009} propose a latent variable design that allows modeling continuous and discrete variables jointly. Another alternative is using Structural Equation Models [SEMs.] While very popular, SEMs were not designed for dichotomous outcomes. \textcite{Bai2020} provides an extensive literature review for multi-outcome models using latent variables.

Bayesian networks [BNs] measure associations between outcome variables as the centerpiece of the model. BNs, however, are designed to depict causal relations in directed acyclic graphs, which constrains the types of associations and dynamics that can be assessed. Yet, most of the methods above set the correlation matrix as a secondary component generally to control for that correlation rather than as a primary outcome, which could rather be the center of the study. 

Looking beyond traditional methods, one alternative to analyze multiple binary outcomes is mapping the problem from a bipartite graph to its line graph representation \parencite{Harary1960}. By focusing on the line graph, the problem then becomes learning the network structure. The Ising model \parencite[see][]{Ising1925, Finnemann2021}, popular in psychometrics and a predecessor of ERGMs \parencite{Park2022}, is used to learn network structure between outcomes. While its mathematical formulation is very close to ERGMs, the Ising model is most commonly used to look at pairs of binary outcomes (dyads.) As the number of outcomes increases, Ising models tend to have a large pool of parameters, and thus, can be awkward to fit and interpret. Two modern alternatives addressing the curse of dimensionality are the eLASSO \parencite{vanBorkulo2015} and a Bayesian approach for Ising models presented in \textcite{Park2022}. While these models can be adapted to fit multiple binary outcomes, they are still limited to dyadic associations, leaving high-order effects (as those illustrated in Figure \ref{fig:dags}) out of the table.

\subsection{Longitudinal studies}

\begin{figure*}
    \centering
    \begin{minipage}[m]{.48\textwidth}
		\large
		\begin{equation*}
			\begin{array}{r}
				A: \linevec{0}{0}\\
				B: \linevec{0}{1}\\
				C: \linevec{1}{0}\\
				D: \linevec{1}{1}
			\end{array} \left(\begin{array}{cccc}
				p_{AA} & p_{AB} & p_{AC} & p_{AD}\\
				p_{BA} & p_{BB} & p_{BC} & p_{BD}\\
				p_{CA} & p_{CB} & p_{CC} & p_{CD}\\
				p_{DA} & p_{DB} & p_{DC} & p_{DD}\\		
			\end{array}\right)
		\end{equation*}
	\end{minipage}\hfill %
	\begin{minipage}[m]{.48\textwidth}
		\centering
		\includegraphics[width=.8\textwidth]{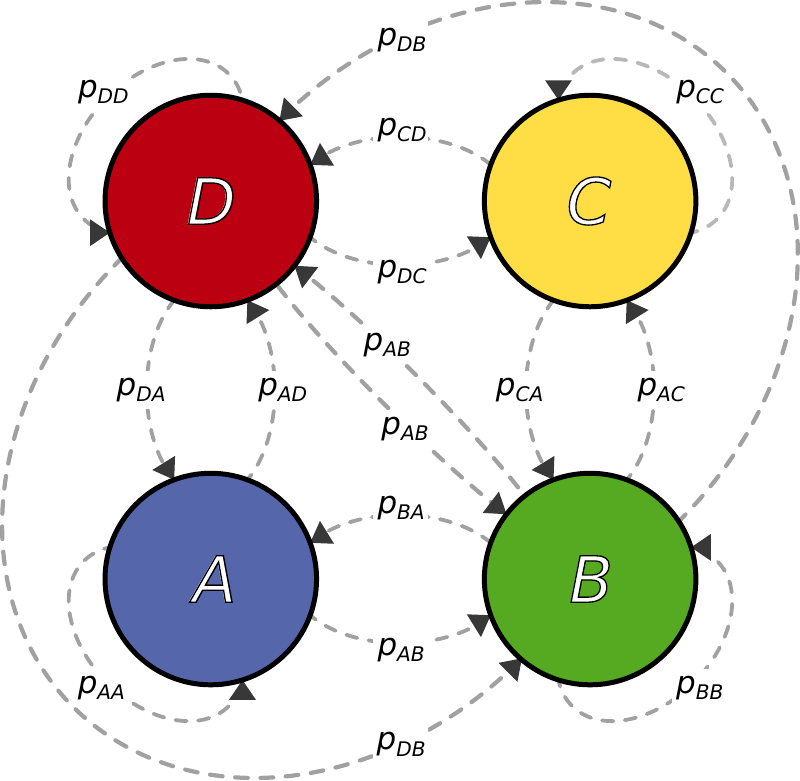}
	\end{minipage}
    \caption{With binary vectors, a Markov process can be characterized by the transition matrix with $2^{2k}$ entries. For example, there are four possible states when dealing with two outcomes. In a Markov order 1 model, these four possible states lead to a four-by-four transition matrix, adding up to 16 parameters.}
    \label{fig:markovmat}
\end{figure*}

Markov Models [MMs] are an appropriate way to analyze longitudinal dichotomous outcomes. Although MMs provide a complete way of capturing the intricacies of complex systems, most of the time, when looking at multi-state systems, the curse of dimensionality constrains the size of the systems MMs can analyze. For example, \textcite{Engelhardt2005} uses Markov models to describe the evolution of gene functions. Their model represents genes as vectors of states indicating the absence or presence of a genetic function. In that context, the Markov transition matrix--which characterizes the stationary state of the system--contains $2^{2K}$ elements, meaning that if one studied ten functions, the number of parameters to estimate would be $2^{2 x 10} \approx 1$ million--See \autoref{fig:markovmat}. Although we can reduce the size of the support by removing transitions in which a gene experiences more than several changes--as in \textcite{Engelhardt2011} where genes were restricted to only two changes,--the interpretation of the estimates is still awkward since parameter estimates may not reflect biological features of the system.

Instead, we proposed using DEFMs to model multi-outcome data. The benefits, as we will show, are that parameter interpretation becomes more manageable, and computational complexity diminishes significantly. The following section describes in detail the formals of the model.

\section{Methods}

\subsection{Setup}

Discrete Exponential-Family Models [DEFMs] have been widely studied in social science. Today, with a large body of scientific literature, Exponential-Family Random Graph Models [ERGMs] \parencite[][and others]{Robins2007, Holland1981, Frank1986, Wasserman1996, Snijders2006} have served as one of the leading platforms by which our understanding of DEFMs has grown. Surprisingly, applications of DEFMs outside of social network analysis are hard to find. We propose using DEFMs to model multiple correlated binary outcomes. 

We are interested in understanding the factors that govern the state of an array of size $K$, $\yvec{it}\equiv\{\y{itk}\} \in \ycal{K}$, with $i$ and $t$ indexing an individual and a point in time, respectively. Now, we say the state of $\yvec{it}$ follows an $m$-order Discrete Markov Process, formally, 

\begin{equation*}
\yvec{it} \sim \mbox{DEFM}\left(\sstat{\yvec{it}, \yvec{it-1},\dots,\yvec{it-m}}, \theta\right),
\end{equation*}

\noindent $t \in \left\{2,\dots,T\right\}$, where $\sstat{\yvec{}}: \ycal{K(m + 1)}\mapsto \mathbb{R}^d$, $d\leq K(m + 1)$, is a function that returns a vector of sufficient statistics, and $\theta\in\Theta$ is a vector of parameters of the same length, $d$. For simplicity, in what follows, we will only consider the case when $m = 1$. In such a case, the probability mass function of this model is

\begin{equation}
\begin{split} 
    \Prcond{\yvec{it}}{\sstat{\cdot},\theta} = & \\ 
    &\hspace{-2cm} \exp{\tp{\theta}\sstat{\yvec{it}, \yvec{it-1}}}\times\normconst{\theta}^{-1},\quad \\ \forall i, t>1
\end{split}
\end{equation}

\noindent where $\normconst{\theta} \equiv \sum_{\yvec{it}'} \exp{\tp{\theta}\sstat{\yvec{it}', \yvec{it-1}}}$ is the normalizing constant. Then, given the Markov Property, the joint distribution of the set $\yvec{} \equiv \{\yvec{i1}, \yvec{i2},\dots,\yvec{iT}\}$ can be computed as follows:

\begin{equation}
\begin{split}
    \Prcond{\yvec{}}{\sstat{\cdot},\theta} & = \\
    & \hspace{-2cm} \left[\prod_{i,t>1}\exp{\tp{\theta}\sstat{\yvec{it}, \yvec{it-1}}}\right]\times \\
    & \hspace{-2cm} \left[\prod_{i}\Prcond{\yvec{i1}}{\theta}\right]\times\normconst{\theta}^{NT}
\end{split}
\end{equation}
\noindent with $\Prcond{\yvec{i1}}{\theta}$ unknown. Observe that in the case of  cross-section data, we can still use this formalism to describe the joint distribution of multiple binary outcomes. Figure \ref{fig:event-in-sns} provides a visual representation of the type of data suited for this model.

\newcommand{\tilt}[1]{\mbox{\small\begin{rotate}{45}{#1}\end{rotate}}}

\begin{figure}
\centering
\begin{tabular}{*{2}{p{.4\linewidth}<\centering}}
\textbf{(A)} Data structure $T$ time points and $K$ outcomes & \textbf{(B)} Example with three outcomes and two time points \\ \\
$ %
\begin{aligned}
     & \qquad\mbox{Outcomes }\rightarrow \\
    \mbox{Time }\downarrow & \left[\begin{array}{cccc} %
         y_{1,1} & y_{1,2} & \dots & y_{1,K}  \\ %
         y_{2,1} & y_{2,2} & \dots & y_{2,K} \\
         \vdots & \vdots & \ddots & \vdots \\
         y_{T,1} & y_{T,2} & \dots &y_{T,K}
    \end{array}\right]
\end{aligned} %
$ & %
$ %
\begin{aligned}
     & 
    \begin{array}{ccc} %
         \hphantom{1} & \hphantom{1} & \hphantom{0} \\%
         \tilt{Alcohol} & \tilt{Tobacco}  & \tilt{Marijuana} %
    \end{array} \\
    \begin{array}{r}
         t = 1 \\
         t = 2 
    \end{array}& \left[\begin{array}{ccc} %
         0 & 1 & 0  \\ %
         1 & 1 & 0 
    \end{array}\right]
\end{aligned} %
$ \\\\
Each row represents a time point and each column a particular outcome. & %
This case represents the transition from only consuming tobacco to consuming tobacco and alcohol.
\end{tabular}
\caption{\textit{Data structure in DEFMs.} Data is represented as binary matrices. In principle, these can be thought of as bipartite graphs.}
\label{fig:event-in-sns}
\end{figure}

In practice, the vector of sufficient statistics will contain terms involving a single outcome and others involving two or more; we call the latter Markov terms. The single outcome terms are, in principle, the covariates that we would use in a Logistic regression, e.g., an intercept term, gender effects, age, etc. The multi-outcome terms--Markov terms--could involve a complex function featuring two or more outcomes. The simplest version of a Markov term is a dichotomous variable equal to one if a given pattern is observed and zero otherwise. For example, we could hypothesize that chewing tobacco, $y_c$ does not happen while smoking tobacco, $y_s$, in such a case, the term would be operationalized as $y_c(1 - y_s) + (1 - y_c)y_s.$

\begin{figure*}
    \begin{equation*}
    \left.\begin{array}{rl}
    L(\theta_1)  = & \logitinv{\theta_1 \sstat{y_1}} \\
    L(\theta_2)  = & \logitinv{\theta_2 \sstat{y_2}} \\
    & \dots \\
    L(\theta_K)  = & \logitinv{\theta_K \sstat{y_K}}
    \end{array}\right\}\mapsto %
    \begin{array}{rl}
    L(\theta_1, \theta_2, \dots, \theta_k, \theta_M) & = \\ %
    & \hspace{-2.5cm}\exp{\sum_k \theta_k \sstat{y_k} + \theta_J \sstat{y_1,\dots, y_K}}\times\normconst{\theta}^{-1}, \\
    \\
    \mbox{where }\theta & =  \left[\theta_1, \dots, \theta_K, \theta_M\right]
    \end{array}
    \end{equation*} 
    \caption{\label{fig:logittodefm}\textbf{Logit and DEFM.} The DEFM is very close to the logistic regression. The only difference is the inclusion of terms that involve two or more outcomes, \textit{e.g.}, an interaction effect. The right-hand side of the diagram shows how a set of independent Logistic models can be combined and extended by incorporating $s(\cdot)$.}
\end{figure*}

\subsection{Model design}

There are as many terms as there are observations in the support function. As we have pointed out before, in the case of Markov Chains of order one, the data in a model with $k$ outcomes has $2^{2k}$ transitions. Small models could include as many as $2^{2k}$ parameters. Nonetheless, such practice is not feasible nor easy to interpret when dealing with more than, for example, three parameters. In a DEFM, we can classify sufficient statistics within the following groups: Non-Markov terms, Transition events, and Association.

\noindent \textbf{A) Non-Markov terms} this class of terms groups all variables involving a single outcome. Furthermore, we can refer to these terms as Logit terms, since we would normally include them in a Logistic regression. Within this class, we only have two types of terms, intercept and interaction effects. The intercept is no more than an indicator variable that equals one if the corresponding outcome equals one. This can be interpreted directly as the logistical regression intercept. When interacting with an exogenous feature like age, the term is akin to the same exogenous variable included in logistic regression. 

\noindent \textbf{B) Transition events} This type of term is related to Markov Chains and may or may not involve more than one outcome. This class of terms is key for evaluating the hypothesis of the sort ``does A leads to B?.'' If the term only involves a single outcome, for example, ``the probability of smoking tobacco given past consumption,'' it can be compared to a lagged outcome term in logistic regression. With multiple outcomes, these terms can be fairly complex, involving activation and deactivation of outcomes over time, or even ignoring the particular state of an outcome at a certain point in time.

\noindent \textbf{C) Association} these terms can be used either in a longitudinal model or in a cross-sectional model. These terms directly capture the interdependence between outcomes. Association terms can capture many classes of interrelatedness; examples include, correlation, one or two-way causation (if longitudinal), and joint distribution. Although some association motifs can also be categorized as transition motifs, these differ in that, by definition, the former involves two or more outcomes. Ultimately, this class of terms is what makes DEFMs special.

\allowdisplaybreaks

\begin{table}[htb]
    \centering
    \begin{tabular}{p{.25\linewidth}<\raggedright p{.275\linewidth}<\raggedright p{.325\linewidth}<\raggedleft}
    \toprule
    & \multicolumn{1}{p{.275\linewidth}<\centering}{\textit{Possible Representation}} &  \multicolumn{1}{p{.325\linewidth}<\centering}{\textit{Obs.} } \\
    \cmidrule(r){2-2}\cmidrule(r){3-3}\multicolumn{2}{l}{\hspace{-.5cm}\textit{A) Non-Markov terms}} \\\\
    A.1) Logit intercept & $\y{a}$ & Baseline prevalence for alcohol consumption \\\\
    A.2) Fixed effect & $\y{a} \times \mbox{Age}$ & Age effect on alcohol \\\\\\
    \multicolumn{2}{l}{\hspace{-.5cm}\textit{B) Transition Events}} \\\\
    B.1) Lagged effect & $(1-\y{a,t-1})\times\y{a,t}$ & Starting alcohol consumption \\\\
    B.2) Causation & $\y{a,t-1}\times(1-\y{s,t-1})\times\y{s,t}$ & Alcohol leads to smoking\\\\\\
    \multicolumn{2}{l}{\hspace{-.5cm}\textit{C) Association terms}} \\\\
    C.1) Correlation & $\y{a}\times\y{s} + (1- \y{a})\times(1- \y{s})$ & Co-ocurrence of alcohol and smoking\\\\
    C.2) One-way causation & $\y{a,t-1}\times(1-\y{s,t-1})\times\y{s,t}$ & Alcohol leads to smoking \\\\
    C.3) Two-way causation & \multicolumn{1}{p{.3\linewidth}}{$\y{a,t-1}\times(1-\y{s,t-1})\times\y{s,t}$\newline $\y{s,t-1}\times(1-\y{a,t-1})\times\y{a,t}$} & Alcohol and smoking lead to each other. \\\\
    C.4) Joint distribution & $\sum_{(\y{a},\y{s}) \in \{0,1\}^2}\y{a}\times\y{s}$ & Full description of joint distribution (all combinations) \\
    \bottomrule
    \end{tabular}
    \caption{\textit{Examples of non-Markov, transition, and association terms (with alcohol and tobacco).} These include at most two outcomes, alcohol, $\y{a}$, and smoking, $\y{s}$. In the case of longitudinal terms, a Markov Chain of order one. In most cases, the variables translate to indicator variables, unless the term is interacting with a continuous covariate like \textit{Age}.}
    \label{tab:example-motifs}
\end{table}

\subsection{Parameter estimation}

Because we can write likelihoods exactly, parameter estimation can be done directly using Maximum Likelihood Estimation. Alternatively, we can also use Bayesian approaches like Maximum A Posteriory [MAP] or Markov-Chain Monte Carlo estimation [MCMC]. Nevertheless, convergence problems proliferate in ERGMs estimation. For example, estimating the effect of triads tends to lead to degenerate models. Recent work has demonstrated that dealing with samples of networks, arrays in our case, reduces estimation problems; moreover, pooled data models using exact likelihoods show higher statistical power and lower type I error rates \parencite{VegaYon2021}.

\subsection{Parameter Interpretation}

This model is closely connected with logistic regression. The method presented here is based on Exponential-Family Random Graph Models, also known as ERGMs. This family of models was designed to study the sufficient statistics--also known as motifs or local structures--that govern the macro properties of networks. For example, hypotheses regarding the prevalence of triangles, high-degree stars, or isolates can be tested using this approach. The fundamental aspect that makes ERGMs appropriate for our problem is the fact that the likelihood function used in these models can be directly used to represent binary arrays. Moreover, DEFMs can be directly mapped to ERGMs of affiliation networks. With that representation, outcomes and individuals would be nodes in the network. A tie from individual $i$ to outcome $k$ in this network would represent a one in our binary array, otherwise, the lack of tie would represent a zero in our binary array.

Parameter interpretation in DEFMs is done with care. Since by construction the model involves identifying motifs that link outcomes, changes in one motif can directly impact another. For example, in a model with three outcomes $\{A, B, C\}$ with motifs capturing $\{A, B\} = \{1,1\}$ and $\{B, C\} = \{1,1\}$, setting $B$ to zero would change both motifs, an thus involving two parameter estimates, $\theta_{AB}$ and $\theta_{BC}$, for posterior analysis. In other words, most of the parameter estimates in DEFMs must be analyzed as one would when looking at interaction effects.

In ERGMs, parameter interpretation is done by looking at what is called change statistics. Specifically, we evaluate changes in odds as a function of perturbations in one entry of the array at a time. Following the previous example, Given that $A$ and $C$ are equal to one, what are the odds between having $B = 0$ and $B = 1$. Formally, the log-odds of $i$ switching the outcome $B$ conditional on the rest of its outcomes, $\yvec{i,-B}$, are:

\begin{equation}
\begin{split}
	\logit{\Prcond{\y{i,B} = 1}{\yvec{i,-B}}} & = \\
	    & \hspace{-2.5cm}\transpose{\theta}\Delta\chng{i,B},
	\end{split}
\end{equation}

\noindent with $\chng{i,B}\equiv \snamed{\yvec{}}{i,B}^+ - \snamed{\yvec{}}{i,B}^-$ as the vector of change statistics, in other words, the difference between the sufficient statistics when $\y{i,B}=1$ and its value when $\y{i,B} = 0$. To show this, we write the following:

\begin{equation*}
\begin{split}
	\Prcond{\y{i,B} = 1}{\yvec{i,-B}} & \\%
		& \hspace{-2.5cm} = \frac{\Pr{\y{i,B} = 1, \yvec{i,-B}}}{%
			\Pr{\y{i,B} = 1, \yvec{i,-B}} + \Pr{\y{i,B} = 0, \yvec{i,-B}}%
			} \\%
		& \hspace{-2.5cm} = \frac{\exp{\transpose{\theta}\s{\yvec{}}^+_{B,i}}}{%
			\exp{\transpose{\theta}\s{\yvec{}}^+_{B,i}} + \exp{\transpose{\theta}\s{\yvec{}}^-_{B,i}} %
			}
\end{split}
\end{equation*}

Therefore, the log-odds of $\Prcond{\y{i,B} = 1}{\yvec{i,-B}}$ is

\begin{equation*}
\begin{split}
& = \log{\frac{\exp{\transpose{\theta}\s{\yvec{}}^+_{B,i}}}{%
		\exp{\transpose{\theta}\s{\yvec{}}^+_{B,i}} + %
		\exp{\transpose{\theta}\s{\yvec{}}^-_{B,i}}}} - \\ %
	& \hspace{1cm} \log{ %
		\frac{\exp{\transpose{\theta}\s{\yvec{}}^-_{B,i}}}{%
			\exp{\transpose{\theta}\s{\yvec{}}^+_{B,i}} + \exp{\transpose{\theta}\s{\yvec{}}^-_{B,i}}}%
	 } \\
 & = \log{\exp{\transpose{\theta}\s{\yvec{}}^+_{B,i}}} - \\
  & \hspace{1cm}\log{\exp{\transpose{\theta}\s{\yvec{}}^-_{B,i}}} \\
 & = \transpose{\theta}\left(\s{\yvec{}}^+_{B,i} - \s{\yvec{}}^-_{B,i}\right) \\
 & = \transpose{\theta}\Delta\chng{i,B}
\end{split}
\end{equation*}
\noindent Henceforth, the conditional probability of changing $B$ to one can be written as:

\begin{equation}\label{eq:gibbs}
\begin{split}
	\Prcond{\y{i,B} = 1}{\yvec{i,-B}} & = \\
	& \hspace{-2.5cm} \frac{1}{1 + \exp{-\transpose{\theta}\Delta\chng{i,B}}},
	\end{split}
\end{equation}
\noindent\textit{i.e.}, a logistic probability. This function is also the Gibbs-sampler, as it describes the full conditional distribution of any given cell in the array.

\subsection{A guide for DEFMs}

The logic of using DEFMs doesn't depart much from either Logistic regression or from ERGMs. A natural way to think about this model is as an enhanced Logistic regression. Furthermore, in the case of not including any term that involves more than one outcome, the $K$ outcome DEFM is then equivalent to $K$ separate logistic regressions (see Figure \ref{fig:logittodefm}). 

\begin{enumerate}
\item \textbf{Motif Census} Although it is recommended to have a predefined set of hypotheses to analyze, it is also important to have a general idea of what motifs are present in the data. To that end, we can always look at simple motif counts in what we call, motif census. This essentially performs a full enumeration of the observed outcome combinations. In cases when there are more than three outcomes, the researcher is recommended to explore the combinations that would make sense. For example, in a study with outcomes $\{A, B, C, D\}$, there are ${\genfrac(){0pt}{1}{4}{3}} = 4$ ways in which we can combine three of them: $\{A, B, C\}$, $\{D, B, C\}$, $\{A, D, C\}$, and $\{A, B, D\}$. For any of those triads, there are $2^3 = 8$ possible motifs, $\{0, 0, 0\}$, $\{1, 0, 0\}$, $\{0, 1, 0\}$, $\{0, 0, 1\}$, $\{1, 1, 0\}$, $\{1, 0, 1\}$, $\{0, 1, 1\}$, and $\{1, 1, 1\}$
    
\item \textbf{Model building} Overall, building DEFMs should be akin to ERGMs. In graph models, we commonly start by using non-Markov terms. In practice, this would mean fitting separate Logistic regressions for each outcome. Once we have a good idea of what non-Markov terms should be included, we can start adding Markov terms to the model. The process of adding terms to the model shouldn't be different from any other MLE approach. Since estimation is done via MLE, the likelihood ratio test to compare models and the evaluation of Markov terms by looking at their significance are natural candidates for assessing model fitness. \textcite{Krivitsky2022} proposes post-estimation analysis techniques for pooled ERGMs, and thus, for DEFMs.
    
\item \textbf{Parameter interpretation} Looking at multiple outcomes means that we will have multiple predicted errors to look at. While we could analyze fitness one outcome at a time, we can still evaluate the model as a whole. Nevertheless, since estimation is done via MLE, the likelihood ratio test to compare models and the evaluation of Markov terms by looking at their significance are natural candidates for assessing model fitness.  In \autoref{sec:example} we provide an example from start to end.
\end{enumerate}

\section{Application: Risk Behaviors in the Social Network Study\label{sec:example}}

Risky health behaviors are a natural candidate for joint modeling. The question of what comes first, \textit{e.g.}, tobacco leads to marijuana, and marijuana leads to other drugs has been a long-standing one. Using DEFMs we can directly test for such transitions while controlling for potential confounding factors. In this section, we illustrate our method with an application using the Social Network Study [SNS] data \parencite{Valente2013,delaHaye2019}.

\subsection{Overview of the SNS}

The SNS data \parencite{Valente2013} is a four-wave survey conducted in Los Angeles county, the United States, that features a sample of 1,795 high-school students. The survey collected information about high-school students between grades 10 to 12, a majority of them self-identified as Hispanic. Among the collected information we have socio-economic status, demographics, social networks, and consumption of alcohol, tobacco, and marijuana--substance use. \textcite{delaHaye2019} focused on tobacco initiation dynamics with an emphasis on social influence (exposure) effects. Here, we will extend their work by incorporating alcohol and marijuana use in the analysis, and assessing how these three interact with each other. Following \textcite{delaHaye2019}, our analysis uses exposure to substance use, ``Hispanic'' (yes/no,) ``Female'' (yes/no,) and ``Academic Grades'' (ranging between 1 to 5, five being the best) as covariates. Restricting the model to individuals who show up at least two times in the dataset, we ended up with 2,028 observations corresponding to 655 individuals. Table \ref{tab:summary-stats} shows summary statistics for the aforementioned variables.

\begin{table}
\centering
\begin{tabular}{l*{4}{c}}
\toprule
& \multicolumn{4}{c}{Wave} \\ \cmidrule(r){2-5}
 & 1 & 2 & 3 & 4\\
Hispanic (yes/no) & 0.64 (0.48) & 0.65 (0.48) & 0.67 (0.47) & 0.68 (0.47)\\
Female (yes/no) & 0.53 (0.50) & 0.56 (0.50) & 0.57 (0.50) & 0.54 (0.50)\\
Academic Grades & 3.94 (0.82) & 3.94 (0.86) & 3.87 (0.80) & 3.92 (0.76)\\\\
\multicolumn{3}{l}{\textit{Substance use (yes/no)}} \\
\quad Alcohol & 0.24 (0.43) & 0.29 (0.45) & 0.41 (0.49) & 0.55 (0.50)\\
\quad Tobacco & 0.03 (0.17) & 0.05 (0.23) & 0.05 (0.22) & 0.21 (0.40)\\
\quad Mj & 0.06 (0.24) & 0.07 (0.25) & 0.11 (0.31) & 0.28 (0.45)\\\\
\multicolumn{3}{l}{\textit{Lagged Exposures (prop. of friends)}} \\
\quad Alcohol & - & 0.28 (0.31) & 0.27 (0.32) & 0.35 (0.37)\\
\quad Tobacco & - & 0.14 (0.23) & 0.10 (0.21) & 0.15 (0.26)\\
\quad Mj & - & 0.13 (0.22) & 0.13 (0.23) & 0.18 (0.28)\\\\
N Obs. & 423 & 437 & 345 & 517\\
\bottomrule
\end{tabular}
\caption{\textit{Summary statistics per wave.} Each column shows the corresponding mean and standard deviation for the variables included in the model. The exposure covariates are not available for the first wave as they were calculated based on the prior wave of substance use behavior. Since friendship networks can change, lagged exposure may increase or decrease over time.}
\label{tab:summary-stats}
\end{table}

\subsection{Motif Census}

As an exploratory step, we start by looking into the motif census. Table \ref{tab:motif-census} shows the motif census for each pair of outcomes. Each entry on the table shows the observed number of observations that match the corresponding patterns. In all cases, the most common motif is the ``empty'' event, in which individuals reported not having used any of the substances in two consecutive time points. 

\begin{table}[tb]
\begin{center}
\begin{minipage}{.3\textwidth}
    \begin{tabular}{l*{4}{p{.1\linewidth}<\centering}}
    \multicolumn{5}{c}{\textit{(A) Alcohol and tobacco}} \\
     & \multicolumn{2}{c}{$t-1$} & \multicolumn{2}{c}{$t$} \\ \cmidrule(r){2-3}\cmidrule(r){4-5}
    count & $y_a$ & $y_t$ & $y_a$ & $y_t$\\
    \midrule
    *639 & 0 & 0 & 0 & 0\\
    *272 & 1 & 0 & 1 & 0\\
    \hphantom{*}110 & 0 & 0 & 1 & 0\\
    \hphantom{*}44 & 1 & 0 & 1 & 1\\
    *33 & 1 & 1 & 1 & 1\\
    \hphantom{*}24 & 0 & 0 & 1 & 1\\
    \hphantom{*}13 & 0 & 0 & 0 & 1\\
    \hphantom{*}11 & 0 & 1 & 1 & 1\\
    *8 & 0 & 1 & 0 & 1\\
    \bottomrule
    \end{tabular}
\end{minipage}\hspace{10pt}
\begin{minipage}{.3\textwidth}
    \begin{tabular}{l*{4}{p{.1\linewidth}<\centering}}
    \multicolumn{5}{l}{\textit{(B) Alcohol and marijuana}} \\
    & \multicolumn{2}{c}{$t-1$} & \multicolumn{2}{c}{$t$} \\ \cmidrule(r){2-3}\cmidrule(r){4-5}
    count & $y_a$ & $y_{mj}$ & $y_a$ & $y_{mj}$\\
    \midrule
    *639 & 0 & 0 & 0 & 0\\
    *225 & 1 & 0 & 1 & 0\\
    \hphantom{*}102 & 0 & 0 & 1 & 0\\
    *67 & 1 & 1 & 1 & 1\\
    \hphantom{*}57 & 1 & 0 & 1 & 1\\
    \hphantom{*}33 & 0 & 0 & 1 & 1\\
    \hphantom{*}14 & 0 & 0 & 0 & 1\\
    \hphantom{*}10 & 0 & 1 & 1 & 1\\
    *7 & 0 & 1 & 0 & 1\\
    \bottomrule
    \end{tabular}
\end{minipage}\hspace{10pt}
\begin{minipage}{.3\textwidth}
    \begin{tabular}{l*{4}{p{.1\linewidth}<\centering}}
    \multicolumn{5}{l}{\textit{(C) Tobacco and marijuana}} \\
    & \multicolumn{2}{c}{$t-1$} & \multicolumn{2}{c}{$t$} \\ \cmidrule(r){2-3}\cmidrule(r){4-5}
    count & $y_t$ & $y_{mj}$ & $y_t$ & $y_{mj}$\\
    \midrule
    *913 & 0 & 0 & 0 & 0\\
    *60 & 0 & 1 & 0 & 1\\
    \hphantom{*}48 & 0 & 0 & 0 & 1\\
    \hphantom{*}36 & 0 & 0 & 1 & 1\\
    \hphantom{*}27 & 0 & 0 & 1 & 0\\
    *26 & 1 & 0 & 1 & 0\\
    \hphantom{*}20 & 1 & 0 & 1 & 1\\
    \hphantom{*}18 & 0 & 1 & 1 & 1\\
    *6 & 1 & 1 & 1 & 1\\
    \bottomrule
    \end{tabular}
\end{minipage}
\end{center}
{\footnotesize(*): Events where there is no change}
\caption{\textit{Motif census.} Each table shows the counts of the observed arrays in the data. In most cases, there is a high prevalence of no change between pairs of outcomes, meaning that changes in substance use were relatively low.}
\label{tab:motif-census}
\end{table}

In \ref{tab:motif-census}.A--alcohol and tobacco,--Most students report no substance use behavior changes as indicated by the rows with asterisks (*), however, there are three transitions worth noting. The most common event in which individuals report a change (110 cases) is reporting alcohol initiation before tobacco. In the second one, counting 44 cases, we find tobacco followed by alcohol as the second most common reported change in the data. And finally, with 11 observations, the least common pattern is individuals who consumed tobacco before consuming alcohol. These three observations provide some evidence that alcohol may be an important precursor to smoking tobacco.

In the case of the pair alcohol and marijuana (\ref{tab:motif-census}.B,) reporting alcohol before ever consuming marijuana was the most common change (102 cases). Similar to the dynamics between alcohol and tobacco, the second most common transition was using marijuana after first using alcohol. And finally, with only 10 cases, reporting marijuana before alcohol was the least common pattern in the data.

With respect to the pair tobacco and marijuana (\ref{tab:motif-census}.C,) the data suggests that individuals will start by tobacco before consuming marijuana. First using tobacco (48 cases,) was the most common change, followed by consuming marijuana after ever consuming tobacco (36 cases.) Similar to what we see with the pairs $\{alcohol, tobacco\}$ and $\{alcohol, marijuana\}$, marijuana as the entry to tobacco consumption was the least frequently reported change in the data. Together, the three tables provide some evidence for the idea that alcohol can be considered a gatekeeper to initiating the consumption of other substances, and what's more, in sequential order. To further explore this, we now proceed to build our model.

\subsection{Model building}

We will study the initiation dynamics in substance use. Our outcome variables will be whether the students had ever drunk alcohol, smoked tobacco, or smoked marijuana. In the context of DEFMs, we are looking at a first-order Markov process in which the initiation of any of these behaviors is a function of both individual features and corresponding past behavior, in other words, whether an individual initiates on tobacco consumption in period $t$ depends on his consumption on alcohol and marijuana in period $t-1$. To represent transition from state A to state B, we write $\left(A\right)\to\left(B\right)$. Furthermore, to represent the transition from zero to one in outcome $k$, we write $(y_k^-)\to(y_k^+)$, with the minus and plus superscripts indicating the variable equals zero or one respectively.

We can reduce the sample space of transitions excluding those that either have little chance of occurring or, as in our case, cannot happen. Since the behaviors we are modeling are ever use of the substance, transitions from one to zero (e. g., ever drinking to never drinking), were excluded from the support set. The particular transitions we are interested in are looking at alcohol and tobacco as gatekeepers for initiating these behaviors. To do this, we include five different motifs in our model: (a) alcohol leading to tobacco, (b) alcohol leading to marijuana, (c) tobacco leading to alcohol, (d) tobacco leading to marijuana, and finally, (e) initiating through marijuana. Table \ref{tab:counts-sns} lists the five motifs, including their mathematical representation and observed counts in the data. Because over half of the population self-identifies as Hispanic, we also include an interaction effect for the first four motifs.

\begin{landscape}

\begin{table}
\begin{center}
\begin{tabular}{m{.005\linewidth} m{.175\linewidth}m{.35\linewidth}<\centering m{.25\linewidth} rr}
\toprule
& Effect & Symbol & Math Rep. & Count & Prop \\ \midrule
\multicolumn{4}{l}{\textit{Pairwise motifs}} \\
& (a) Alcohol leading to\newline \hphantom{to} tobacco& $(alcohol^+, \underline{tobacco^-})$ $\to$ $(alcohol^+, \underline{tobacco^+})$ & $y_{t-1,a}(1-y_{t-1,s})y_{t,a}y_{t,s}$ & 44 & 3.81\%\\
& (b) Alcohol leading to\newline \hphantom{to} marijuana & $(alcohol^+, \underline{mj^-})$ $\to$ $(alcohol^+, \underline{mj^+})$ & $y_{t-1,a}(1-y_{t-1,mj})y_{t,a}y_{t,mj}$ & 57 & 4.94\%\\
& (c) Tobacco leading to\newline \hphantom{to} alcohol & $(\underline{alcohol^-}, tobacco^+)$ $\to$ $(\underline{alcohol^+}, tobacco^+)$ & $y_{t-1,s}(1-y_{t-1,a})y_{t,t}y_{t,a}$ & 11 & 0.95\%\\
& (d) Tobacco leading to\newline \hphantom{to} marijuana & $(tobacco^+, \underline{mj^-})$ $\to$ $(tobacco^+, \underline{mj^+})$ & $y_{t-1,s}(1-y_{t-1,mj})y_{t,s}y_{t,mj}$ & 20 & 1.73\%\\\\
\multicolumn{4}{l}{\textit{Pairwise motifs mediated by race}} \\
& & $(alcohol^+, tobacco^-)$ $\to$ $(alcohol^+, tobacco^+)$ \newline  x~Hispanic & & 40 & 3.47\%\\
& Same as before but \newline \hphantom{to} $\times$ Hispanic & $(alcohol^+, mj^-) \to (alcohol^+, mj^+)$ \newline x~Hispanic & Same as before but \newline \hphantom{to} $\times$ Hispanic & 51 & 4.42\%\\
& & $(alcohol^-, tobacco^+)$ $\to$ $(alcohol^+, tobacco^+)$ \newline  x~Hispanic & & 10 & 0.87\%\\\\
\multicolumn{4}{l}{\textit{Motif involving all three outcomes}} \\
& (e) Marijuana before\newline \hphantom{to} tobacco and alcohol& $(alcohol^-, tobacco^-, \underline{mj^-})$ $\to$ \newline $\hphantom{\to}(alcohol^-, tobacco^-, \underline{mj^+})$ & $(1-y_{t-1,a})(1-y_{t-1,s})(1-y_{t-1,mj})$\newline $\hphantom{\to}(1-y_{t,a})(1-y_{t,s})y_{t,mj}$ & 10 & 0.87\%\\
\bottomrule
\end{tabular}
\caption{\textit{Motifs in the SNS dataset.} The table shows the effects (motifs) used, corresponding representation (symbol,) observed counts, and relative proportions in the dataset. All the motifs used here are transition motifs. Motifs may interact with other covariates. In our analysis, the three pairwise motifs studied were also incorporated interacting with ``Hispanic.''}
\label{tab:counts-sns}
\end{center}
\end{table}

\end{landscape}

Data processing was conducted in R version 4.2.1 \parencite{R}, with the R packages data.table \parencite{Dowle2021}, netdiffuseR \parencite{Valente2020,netdiffuseR}, and texreg \parencite{Leifeld2013}. Model fitting was done using the defm R package, which we developed, and is available at \url{https://github.com/UofUEpi/defm}.

\begin{table}
\begin{center}
    \begin{tabular}{p{.65\linewidth} D{)}{)}{9)3}}
\toprule
\textit{alcohol}                                                                    &                       \\
\quad (Intercept)                                                          & -2.83 \; (0.66)^{***} \\
\quad  x Hispanic                                                          & 0.89 \; (0.24)^{***}  \\
\quad  x Female                                                            & 0.18 \; (0.20)        \\
\quad  x Academic Grades                                                   & -0.08 \; (0.14)       \\
\quad  x Expo. Drink                                                       & -0.33 \; (0.30)       \\
\textit{tobacco}                                                                    &                       \\
\quad (Intercept)                                                          & -3.97 \; (0.86)^{***} \\
\quad  x Hispanic                                                          & 0.48 \; (0.41)        \\
\quad  x Female                                                            & -0.73 \; (0.26)^{**}  \\
\quad  x Sib. Smokes                                                       & 1.02 \; (0.30)^{***}  \\
\quad  x Academic Grades                                                   & -0.16 \; (0.17)       \\
\quad  x Expo. Smoke                                                       & 1.61 \; (0.42)^{***}  \\
\textit{mj}                                                                         &                       \\
\quad (Intercept)                                                          & -1.89 \; (0.84)^{*}   \\
\quad  x Hispanic                                                          & 0.58 \; (0.41)        \\
\quad  x Female                                                            & -0.66 \; (0.24)^{**}  \\
\quad  x Academic Grades                                                   & -0.33 \; (0.16)^{*}   \\
\quad  x Expo. MJ                                                          & 0.44 \; (0.42)        \\\\
\textit{Pairwise Motifs}                                                   &                       \\
\quad (a) $(alcohol^+, tobacco^-) \to (alcohol^+, tobacco^+)$             & 1.12 \; (0.66)        \\
\quad (b) $(alcohol^+, mj^-) \to (alcohol^+, mj^+)$                       & -0.09 \; (0.60)       \\
\quad (c) $(alcohol^-, tobacco^+) \to (alcohol^+, tobacco^+)$             & 1.27 \; (1.37)        \\
\quad (d) $(tobacco^+, mj^-) \to (tobacco^+, mj^+)$                       & 1.03 \; (0.36)^{**}   \\\\
\textit{Pairwise Motifs by ethnicity}                                &                       \\
\quad \hphantom{(a)}$(alcohol^+, tobacco^-) \to (alcohol^+, tobacco^+)$ x Hispanic  & -0.24 \; (0.72)       \\
\quad \hphantom{(a)}$(alcohol^+, mj^-) \to (alcohol^+, mj^+)$ x Hispanic            & 0.00 \; (0.64)        \\
\quad \hphantom{(a)}$(alcohol^-, tobacco^+) \to (alcohol^+, tobacco^+)$ x Hispanic  & 0.61 \; (1.47)        \\\\
\textit{Motif involving all three outcomes}                          &                       \\
\quad (e) $(alcohol^-, tobacco^-, mj^-) \to (alcohol^-, tobacco^-, mj^+)$ & -2.17 \; (0.38)^{***} \\
\midrule
AIC                                                                        & 1679.24               \\
BIC                                                                        & 1830.77               \\
N                                                                          & 1722 \\
N events (transitions)                                                     & 1154                  \\
\bottomrule
\multicolumn{2}{l}{\scriptsize{$^{***}p<0.001$; $^{**}p<0.01$; $^{*}p<0.05$}.}
\end{tabular}
\caption{\textit{Substance use initiation dynamics}. The table shows coefficients and standard errors in parentheses. The modeling fitting was done using maximum likelihood estimation.}
\label{table:snsmodelfit}
\end{center}
\end{table}

\subsection{Analysis}

The table \ref{table:snsmodelfit} shows the results of the final model. Regarding the non-Markov effects, the Hispanic population is significantly more likely to have ever consumed alcohol (estimate of 0.89). Females are less likely to report using tobacco or marijuana, with estimates of -0.73 and -0.66 respectively, but with no difference in alcohol consumption. Reporting higher academic grades is negatively associated with marijuana initiation (estimate of -0.33,) but not with the other two outcomes. In the case of tobacco initiation, having a sibling who smokes (estimate of 1.02) is a strong indicator of tobacco initiation. Finally, regarding social influence, our model suggests that only smoking tobacco may be influenced by exposure to friends who reported ever smoking tobacco.

In the case of the Markov effects, of the eight motifs included in our model, only two are relevant. First, tobacco leading to marijuana (motif d) shows an important association, with an estimate of 1.03. Secondly, initiating through marijuana, in other words, with an estimate of -2.17, reporting marijuana as the first substance used (motif e) is highly unlikely. None of the pairwise motifs involving tobacco resulted in a significant effect.

Finally, our results for tobacco initiation dynamics are in line with \textcite{delaHaye2019}, except for academic grades, which they reported to be significant.

\section{Discussion}

In this paper, we proposed a new method for studying multiple jointly-distributed binary outcomes. Our method, which we call the Discrete Exponential-Family Model [DEFM], borrows from the long tradition of Exponential-Family Random Graph Models [ERGMs.] While methods for estimating jointly-distributed outcomes exist, most of those have the primary goal of examining inferences despite the interrelatedness. In our method, doing inferences about associations between the outcomes is the primary objective. 

A central piece of our model is sufficiency. By building hypotheses based on  sufficient statistics/motifs, we can avoid the curse of dimensionality, allowing us to build models featuring a large number of outcomes in a computationally efficient way. Moreover, using this framework, we can model complex associations between outcomes, from the simple correlation between two outcomes to highly non-linear associations involving three or more outcomes. Since our model builds on ERGMs, all the theory developed around it is readily available to DEFMs, including Bayesian methods, estimating models with missing data, count data, etc.

The example presented here of adolescent progression through multiple substance use trajectories is one that is reasonably prominent in public and adolescent health research (). Using our model, we were able to identify tobacco consumption as an important predecessor to marijuana consumption. Furthermore, our model suggests that there's little chance individuals start substance use through marijuana. There are numerous other applications of this methodology both in clinical and population settings. For example, well-known Knowledge, Attitude, and Practice (KAP) surveys could be analyzed to determine whether the KAP sequence of steps occurs, or one of the other five combinations, KPA, AKP, APK, PKA, or PAK \parencite{VALENTE1998}. Or analyses of policy and behavior diffusion progressing through the knowledge, persuasion, decision, trial, and implementation stages \parencite{Rogers2003, Valente2019}. In clinical research, one could study the progression of various diseases such as diabetes leading to neuropathy leading to cardiovascular disease or traumatic brain injury leading to sequelae such as cognitive impairment or dementia. Finally, one can also apply this methodology to evaluation research by specifying intervention exposure as one of the nodes leading to positive outcomes or characteristics associated with being exposed to an intervention.

Finally, Discrete-Family Exponential Models, including estimation, simulation, and post-estimation analyses, are readily available to practitioners through our R package, \texttt{defm}.

\printbibliography

\pagebreak

\appendix

\section{Acknowledgment}

This work was supported by the Assistant Secretary of Defense for Health Affairs and endorsed by the Department of Defense, through the Psychological Health/Traumatic Brain Injury Research Program Long-Term Impact of Military-Relevant Brain Injury Consortium (LIMBIC) Award/ W81XWH-18-PH/TBIRP-LIMBIC under Award No. I01 RX003443. The US Army Medical Research Acquisition Activity, 839 Chandler Street, Fort Detrick MD 21702-5014 is the awarding and administering acquisition office. Dr. Pugh was also supported by VA
Health Services Research and Development Service Research Career Scientist Award, 1 IK6 HX002608. Opinions, interpretations, conclusions, and recommendations are those of the author and are not necessarily endorsed by the Department of Defense or the Department of Veterans Affairs. Any opinions, findings, conclusions, or recommendations expressed in this publication are those of the author(s) and do not necessarily reflect the views of the US Government, the US Department of Veterans Affairs, or the Department of Defense and no official endorsement should be inferred.





\end{document}